\documentclass[12pt]{iopart}
\usepackage{graphicx}
\usepackage{cite}

\begin{document}

\title[Formation and stimulated photodissociation]
{Formation and stimulated photodissociation of metastable molecules with emission of photon at the collision of two atoms in a laser radiation field.}

\author{E. Gazazyan, A. Gazazyan}

\address{Institute for Physical Research, NAS of Armenia, Ashtarak-2, 0203, Armenia}
\ead{emilgazazyan@gmail.com}
\vspace{10pt}
\begin{indented}
\item[]November 2016
\end{indented}

\begin{abstract}
The Formation of metastable molecules (Feshbach resonances) at the collision of two atoms and subsequent stimulated transition to a lower unbound electronic molecular state, with emission of a photon of the laser radiation has been investigated. This can develop, in particular, for $Rb_2$ molecules due to resonance scattering of two $Rb$ atoms. The considered process is a basis for the creation of excimer lasers. Expressions for the cross sections of elastic and inelastic resonance scattering and the intensity of the stimulated emission of the photons have been obtained. 
\end{abstract}

%
%
%
%
%

\section{Introduction}

Metastable molecules are formed due to collision of two atoms, when the energy of a bound molecular state in a closed channel is close to the energy of two atoms in the center of mass in an open channel. The weak coupling between the channels leads to the strong mixing of them, and the resulting metastable molecular states, which called the Feshbach resonances \cite{1}, will have a finite lifetime and will decay both in the initial channel and in the other channels. The process of formation if intermediate metastable molecular states can be controlled, in particular, by the external magnetic field by the change of the resonance detuning, when the bound state in a closed channel has the different hyperfine state then the incoming atoms in the open channel. In this case, the difference between the hyperfine states, which is caused by the difference in the Zeeman shifts, can be controlled by the detuning of the Feshbach resonance by means of the magnetic field \cite{2,3,4,5}. In case where magnetically tunable resonances are absent, an alternative method for control of Feshbach resonance is the optical method [5] (optical Feshbach resonance). Optical control of Feshbach resonances by means of quantum interference was proposed in work \cite{6,7,8}.

In the theory of resonant collision, in addition to Feshbach method \cite{1}, which has been developed for studies of nuclear reactions and is successfully applied for collision atoms in BEC, there is an alternative Fano approach \cite{9} exists exploiting the configuration interaction in multielectron atoms. Both approaches assume appearance of resonance phenomena when discrete states are coupled with continuum. Fano technique is usually associated with asymmetry of shapes of resonance lines which is known in atomic physics as “Fano profile”. Similar interference phenomena of asymmetry of resonance line shape are also observed in nuclear reactions \cite{10}. Fano technique is, however, used not only in atomic physics for, e.g., studies of autoionization and Rydberg states [9], resonance ionization of atoms \cite{11,12}, and laser induced continuum structures (LICS) \cite{13}. The Fano technique is also used for considerations of resonant collision [9] including those of electrons with atoms with formation of negative ions and interference phenomena in the field of laser radiation \cite{14,15}. Fano technique is widely used in also other fields of physics. It is, e.g., used for explanation of asymmetry in the absorption impurity ions in crystals, which is caused by for formation of excitonic resonances \cite{16,17}. With use of these resonances works \cite{18,19}, study the phenomenon of storage and reconstruction of quantum information in solids. The asymmetric form of the Fano resonance important in considerations of nanoscale structures of interacting  quantum systems \cite{20,21}.     

In the present work we consider collision of atoms with formation of the metastable molecules (Feshbach resonances) and the subsequent stimulated transition to the lower unbound molecular electronic states with emission of photon of the laser radiation. Expressions for the cross sections of the elastic and inelastic resonance scattering and the intensity of the stimulated emission of the photons have been obtained.

\section{Formation of Feshbach Resonance}

Consider elastic and inelastic collision of two atoms with formation of Feshbach resonance (Fig.1). $U$ the interaction which couples electronic states in the open and closed channel. Laser radiation with frequency $\omega$ couples the upper molecular quasi bound state with the lower uncoupled molecular state with interaction $\Omega_E$.
\begin{figure}[h]
	\centering
	\includegraphics[width=0.6\linewidth]{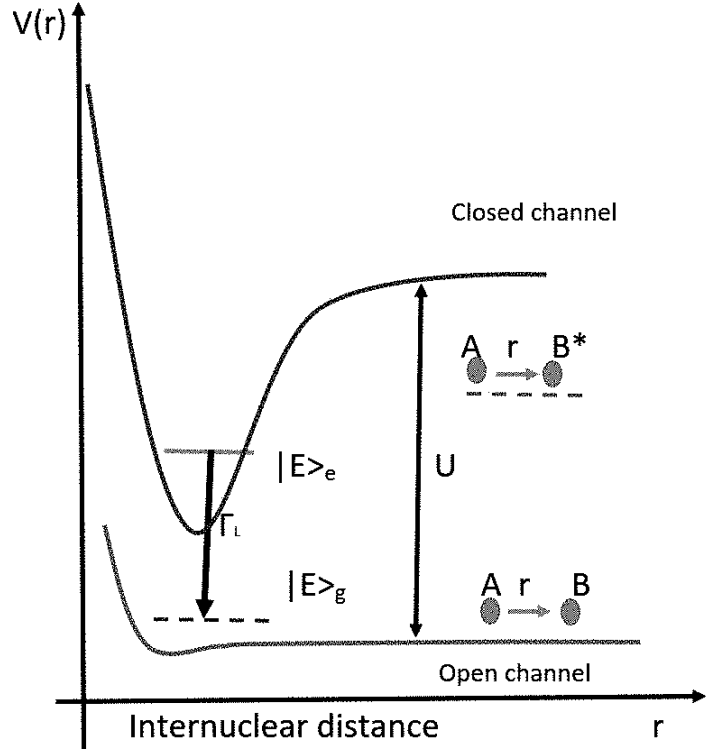}
	\caption{Diagram of formation of metastable molecules (Feshbach resonances) at the collision of two atoms and subsequent stimulated transition to a lower unbound electronic molecular state, with emission of a photon of the laser radiation. Here $B^*$ is excited atom B.}
	\label{fig:MolDiag}
\end{figure}
The Hamiltonian for the considered process (Fig.1) has the following form:

\begin{eqnarray}
\label{eq:Hamiltonian}
H=E_e|e\rangle\langle e|+\int dE E\bigg(|E\rangle_{11}\langle E|+|E^{-\omega}\rangle_{22}\langle E^{-\omega}|\bigg)+\nonumber \\ \int dE \bigg(U_e|e\rangle _1\langle E|+U^*_E|E\rangle\langle e|\bigg)+ \nonumber \\ \int dE \bigg(\Omega_{E-\omega}(t)|e\rangle _2\langle E-\omega|+\Omega^*_{E-\omega}(t)|E-\omega\rangle_2\langle e|\bigg)   
\end{eqnarray}
We represent the solution of the Schr\"{o}dinger equation with the Hamiltonian (\ref{eq:Hamiltonian}) as:
\begin{equation}\label{eq:Shrodinger}
|\Psi(t)\rangle = C_e(t)|e\rangle e^{-iE_et}+ \int dE e^{-iEt}(b_{1E}(t)|E\rangle _1+e^{i\omega t}b_{2,E-\omega}(t)|E-\omega\rangle _2)
\end{equation}
Substituting expression (\ref{eq:Shrodinger}) 
for the wave vector and using that $\Omega(t)=\Omega e^{-i\omega t}$, where $\omega$ is the frequency of laser radiation, into the Shr\"{o}dinger equation we obtain the following system of differential equations for the coefficients of the expansion (\ref{eq:Shrodinger}): 

\begin{eqnarray}
i\frac{dC_e(t)}{dt}=\int dE e^{i(E_e-E)t}(b_{1,E}(t)U_E+b_{2,E-\omega}(t)\Omega_{E-\omega}) \nonumber \\
i\frac{db_{1,E}(t)}{dt}=c_e(t)e^{-i(E_e-E)t}U^*_E\\
i\frac{db_{2,E-\omega}(t)}{dt}=c_e(t)e^{-i(E_e-E)t}\Omega^*_{E-\omega}\nonumber
\end{eqnarray}

After the Fourier transform for the coefficient in the formula (\ref{eq:Shrodinger})
\begin{eqnarray}\label{eq_4:Coefficient}
C_e(t)=\int d\lambda e^{-i(\lambda-E_e)t}C_E(\lambda) \nonumber \\
b_{1,E}(t)=\int d\lambda e^{-i(\lambda-E)t}b_{1,E}(\lambda) \\
b_{2,E-\omega}(t)=\int d\lambda e^{-i(\lambda-E)t}b_{2,E-\omega}(\lambda)\nonumber 
\end{eqnarray}
for the Fourier components we obtain for the Fourier components of the expansion coefficient the following system equations:
	
	\begin{eqnarray}\label{eq:5a}
	(\lambda-E_e)C_e(\lambda)=\int dE(U_Eb_{1,E}(\lambda)+\Omega_{E-\omega}b_{2,E-\omega}(\lambda)) 
	\end{eqnarray}
	\begin{equation}\label{eq:5b}
	(\lambda -E)b_{1,E}(\lambda)=U^*_EC_e(\lambda) 
	\end{equation}
	\begin{equation}\label{eq:5c}
	(\lambda -E )b_{2,E-\omega}(\lambda)=\Omega^*_{E-\omega}C_e(\lambda)
	\end{equation}

Now we obtain from the equation (\ref{eq:5b}),(\ref{eq:5c}) \cite{10}:

\begin{eqnarray}\label{eq:8}
b_{1,E}(\lambda)=[\frac{P}{\lambda-E}+Z(\lambda)\delta(\lambda-E)]U^*_EC_e(\lambda)
\end{eqnarray}
\begin{eqnarray}\label{eq:9}
b_{2,E-\omega}(\lambda)=[\frac{P}{\lambda-E}+Z(\lambda)\delta(\lambda-E)]\Omega^*_{E-\omega}C_e(\lambda),
\end{eqnarray}

where

\begin{equation}\label{eq7}
Z(\lambda)=\frac{\lambda-E_e-\Delta(\lambda)}{\Gamma(\lambda)}
\end{equation}
In expression (\ref{eq7}) $\Delta(\lambda)$ and $\Gamma(\lambda)$ are the full resonant shifts and width and $P$ denotes the principle value:
\begin{eqnarray} 	
	\Delta(\lambda)=\Delta_F(\lambda)+\Delta_L(\lambda) \nonumber \\
	\Delta_F(\label)=P\int dE\frac{|U_E|^2}{\lambda-E} \nonumber \\
	\Delta_L(\label)=P\int dE\frac{|\Omega_{E-\omega}|^2}{\lambda-E}  \nonumber\\
	\Gamma(\lambda)=\Gamma_F(\lambda)+\Gamma_L(\lambda) \nonumber\\
	\Gamma_F(\lambda)=2\pi|U_{\lambda}|^2 \nonumber\\
	\Gamma_L(\lambda)=2\pi |\Omega_{\lambda-\omega}|^2 \nonumber
\end{eqnarray} 
By substituting expressions (\ref{eq:8}),(\ref{eq:9}) into formula (\ref{eq:Shrodinger}) from the ortonormalization condition we obtain the following expression for the $C_e(\lambda)$:
\begin{equation}\label{eq8}
C_e(\lambda)=\sqrt{\frac{2\pi}{\Gamma(\lambda)}\frac{1}{z^2(\lambda)+\pi^2}}
\end{equation} 
For the first solution obtain the following expression:

\begin{eqnarray}\label{eq9}
|\Phi^{(1)}_\lambda (t)\rangle=c_e(\lambda) \bigg[ |e\rangle +\int dE \bigg( \frac{P}{\lambda-E}+z(\lambda)\delta(\lambda-E) \bigg)*\nonumber \\ \bigg(U^*_E|E\rangle _1 +e^{i\omega t}\Omega^*_{E-\omega}|E-\omega \rangle _2 \bigg) \bigg]
\end{eqnarray}

For the second ortonormalized solution in the case of $c_e(\lambda)=0$,  we have the following expression 
\begin{equation}\label{eq10}
|\Phi^{(2)}_\lambda (t)\rangle =\sqrt{\frac{2\pi}{\Gamma (\lambda)}} \bigg(\Omega_{\lambda-\omega}|\lambda\rangle _1 - e^{i\omega t}U_\lambda|\lambda-\omega\rangle _2\bigg)
\end{equation}
This solutions (\ref{eq9}) and (\ref{eq10})  are provide the ortonormalization condition for quasienergy functions:
\begin{equation} \label{eq11}
\langle \Phi^{(j')}_{\lambda '}(t)| \Phi^{(j)}_{\lambda}(t)\rangle=\delta_{j',j}\delta(\lambda '-\lambda)
\end{equation}

\section{Cross sections of elastic and inelastic scattering and intensity of emission radiation.}

Asymptotic ($r\rightarrow \infty $)	expressions for continuous-spectrum wave functions with orbital angular momentum l are known to have the following appearance: 
	\begin{eqnarray}\label{eq12a}
	|E\rangle^{(l)}_1\propto \frac{1}{k_1r_1}\sin\bigg(k_1r_1+\delta_1-\frac{1}{2}\pi l_1\bigg)P_{l_1}(\cos\Theta_1), \hspace{1cm} k_1=k(E) \\	
	|E-\omega\rangle^{(l)}_2\propto \frac{1}{k_2r_2}\sin\bigg(k_2r_2+\delta_2-\frac{1}{2}\pi l_2\bigg)P_{l_2}(\cos\Theta_2), \hspace{0.3cm} k_2=k(E-\omega)
	\end{eqnarray}
with $P_l(\cos\Theta)$ begin the Legendre polynomials.
Taking into account that the wave functions of bound states of atoms vanish asymptotically ($|e\rangle=0$) at large distance ($r\rightarrow \infty $) we can write the quasienergy wave functions (\ref{eq9})(\ref{eq10}) as follows:

\begin{eqnarray}\label{eq13}
|\Phi^{(1)}_\lambda(t)\rangle=-\frac{\pi c_e(\lambda)}{\sin\eta}\bigg(\frac{U^*_\lambda}{k(\lambda)r_1}\sin\bigg(k(\lambda)r_1+\eta+\delta_1-\frac{l_1\pi}{2}\bigg)P_{l_1}(\cos\Theta_1)+\nonumber\\
e^{i\omega t}\frac{\Omega^*_{\lambda-\omega}}{k(\lambda-\omega)r_2}\sin \bigg(k(\lambda-\omega)r_2+\eta+\delta_2-\frac{l_2\pi}{2}\bigg)P_{l_2}(\cos\Theta_2)\bigg)
\end{eqnarray}

\begin{eqnarray}\label{eq14}
|\Phi^{(1)}_\lambda(t)\rangle=\sqrt{\frac{2\pi }{\Gamma(\lambda)}}\bigg(\frac{\Omega_{\lambda-\omega}}{k(\lambda)r_1}\sin\bigg(k(\lambda)r_1+\delta_1-\frac{l_1\pi}{2}\bigg)P_{l_1}(\cos\Theta_1)-\nonumber \\
e^{i\omega t}\frac{u_\lambda}{k(\lambda-\omega)r_2}\sin \bigg(k(\lambda-\omega)r_2+\delta_2-\frac{l_2\pi}{2}\bigg)P_{l_2}(\cos\Theta_2)\bigg)
\end{eqnarray}
where $\delta_l$ is the phase of non-resonant scattering and $\eta$ is the phase caused by resonant Feshbach scattering.
\begin{equation}\label{eq:15}
\tan\eta=-\frac{\pi}{z(\lambda)}
\end{equation}
We now represent the scattering state vector $|\Phi_\lambda(1\rightarrow1,2)\rangle$ at $r\rightarrow \infty$ as superposition of quasienergy function (\ref{eq13}),(\ref{eq14})
\begin{equation}\label{eq16}
|\Phi_\lambda(1\rightarrow1,2)\rangle=\sum_jA_j|\Phi_\lambda^{(j)}(t)\rangle
\end{equation} 
with
\begin{equation}\label{eq17}
\sum_j|A_j|^2=1
\end{equation}
then, if we require the presents of incoming and outgoing waves in the first, elastic, channel and the absence of incoming wave in the second, inelastic, channel we can write for the expansion coefficients in (\ref{eq16}) $A_j$ the following:
\begin{eqnarray}\label{eq18}
A_1=\frac{U_\lambda}{\sqrt{|U_\lambda|^2+|\Omega_{\lambda-\omega}|^2}},\hspace{1cm}  A_2=\frac{\Omega^*_{\lambda-\omega}e^{-i\eta}}{\sqrt{|U_\lambda|^2+|\Omega_{\lambda-\omega}|^2}}
\end{eqnarray}
From expressions (\ref{eq18}) we can obtain for scattering state vector (\ref{eq16}):
\begin{eqnarray}\label{eq19}
|\Phi_\lambda(1\rightarrow 1,2)\rangle=\frac{e^{-i(\delta_1+\eta)}}{k(\lambda)r_1}\bigg[\sin(k(\lambda)r_1-\frac{l_1\pi}{2})-\frac{e^{2i\delta_1}}{2i} \bigg(\frac{\Gamma_F(\lambda)}{\Gamma(\lambda)}\bigg(1-e^{2i\eta }\bigg)+ \nonumber \\
+e^{-2i\delta_1}-1\bigg) e^{i(k(\lambda)r_1-\frac{l_1\pi}{2})}\bigg] P_{l_1}(\cos\Theta_1)-e^{i\omega t} \frac{\sin\eta}{k(\lambda-\omega)r_2} \frac{\sqrt{\Gamma_F(\lambda)\Gamma_L(\lambda)}}{\Gamma(\lambda)} \nonumber \\ e^{i\delta_2}e^{-i(k(\lambda-\omega))r_2-\frac{l_2\pi}{2}}P_{l_2}(\cos\Theta_2)
\end{eqnarray}
From expression (\ref{eq19}) we can obtain the formulas for corresponding cross sections

	\begin{eqnarray}\label{eq20a}
	\sigma(1\rightarrow 1)=\frac{4\pi(2l+1)}{k^2(\lambda)}\frac{1}{4}\bigg|\frac{\Gamma_F(\lambda)}{\Gamma(\lambda)}(1-e^{2i\eta})+e^{-2i\delta_l}-1\bigg|^2\\	
	\sigma(1\rightarrow 2)= \frac{4\pi(2l+1)}{k^2(\lambda-\omega)}\frac{\Gamma_F(\lambda)\Gamma_L(\lambda)}{\Gamma^2(\lambda)}\sin^2\eta
	\end{eqnarray}

Let as separate in (\ref{eq20a}) for elastic scattering the resonant state and write the potential part of scattering cross sections in the form
\begin{equation}\label{eq:22}
\sigma_{pot}=\frac{\pi(2l+1)}{k^2(\lambda)}4sin^2\delta_l
\end{equation}
The total cross section of elastic scattering we write as:
\begin{equation}\label{eq23}
\sigma^{el}_{tot}=\sigma_{pot}+\sigma^{el}_{res}
\end{equation}
where
\begin{equation}
\sigma^{el}_{res}=\frac{\pi(2l+1)}{k^2(\lambda)}\frac{\Gamma^2_F(\lambda)}{\Gamma^2(\lambda)}\bigg|1-e^{2i\eta}\bigg|^2
\end{equation}

From expressions (\ref{eq7}) and (\ref{eq:15}) for the resonant elastic and inelastic cross sections, we obtain:

	\begin{eqnarray}\label{eq25a}
	\sigma^{el}_{res}=\frac{\pi(2l+1)}{4k^2(\lambda)}\frac{\Gamma^2_F(\lambda)}{(\lambda-E_e-\Delta(\lambda))^2+\frac{\Gamma^2(\lambda)}{4}} \\
	\sigma^{inel}_{res}=\frac{\pi(2l+1)}{k^2(\lambda)}\frac{\Gamma_F(\lambda)\Gamma_L(\lambda)}{(\lambda+\omega-E_e-\Delta(\lambda))^2+\frac{\Gamma^2(\lambda)}{4}}
	\end{eqnarray}
	
 \begin{figure}[h]
 	\centering
 	\includegraphics[width=0.6\linewidth]{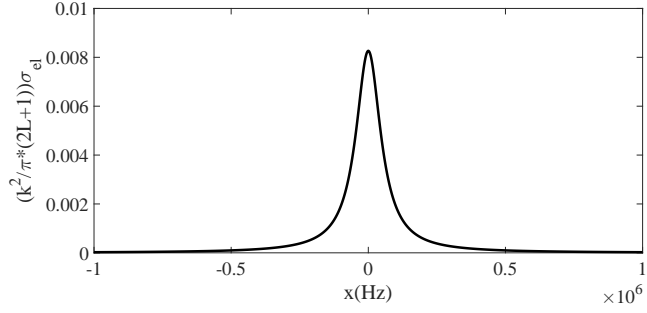}
 	\caption[1]{Cross section for elastic collisions vs detuning ($x=\lambda - E_e-\Delta(\lambda))$}
 	\label{fig:Elastic}
 \end{figure}
 
 \begin{figure}[h]
 	\centering
 	\includegraphics[width=0.6\linewidth]{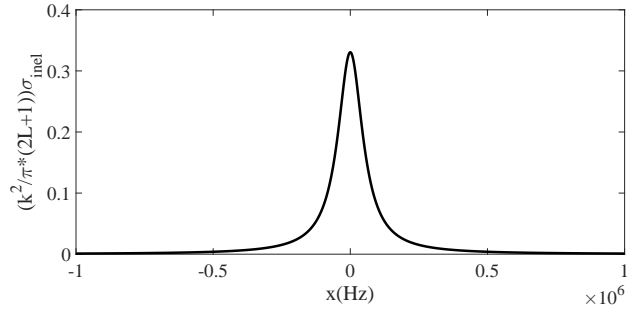}
 	\caption[1]{Cross section for inelastic collisions vs detuning ($x=\lambda +\omega- E_e-\Delta(\lambda))$}
 	\label{fig:Inelastic}
 \end{figure}
Distribution for the spectral intensity of the emission radiation at the stimulated transition to the lower unbound molecular state has the following form \cite{22}:
\begin{equation}\label{eq26}
I(\omega)=I_0\frac{\Gamma}{2\pi}\frac{1}{(\omega+\lambda-E_e-\Delta(\lambda))^2+\frac{\Gamma^2}{4}}
\end{equation}  

Where $I_0$ is a full intensity of incident laser radiation field.
\begin{equation*}
I_0=\int^{\infty}_{-\infty}I(\omega)d\omega
\end{equation*}
The spectral intensity of the stimulated emission of photons coming from unity volume of the gas are obtained:
\begin{equation}\label{eq27}
I(\omega)=N^*I_0\frac{\Gamma}{2\pi}\frac{1}{(\omega+\lambda-E_e-\Delta(\lambda))^2+\frac{\Gamma^2}{4}}
\end{equation}
\begin{figure}[h]
	\centering
	\includegraphics[width=0.6\linewidth]{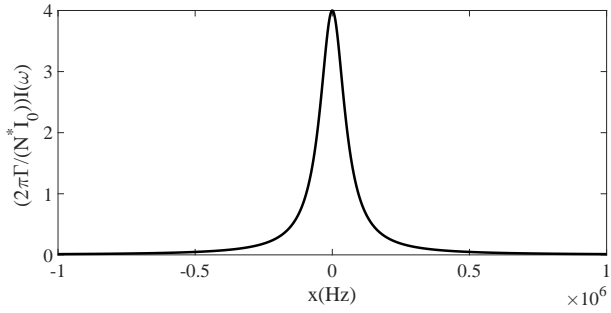}
	\caption[1]{The spectral intensity of the stimulated emission of photons vs detuning ($x=\omega+\lambda - E_e-\Delta(\lambda))$ }
	\label{fig:Inten}
\end{figure}
In figures \ref{fig:Elastic},\ref{fig:Inelastic},\ref{fig:Inten}, shows the curves corresponding to the formulas (29,30,32), for the values $\Gamma_f=10^4Hz$ and $\Gamma_L=10\Gamma_f$, where $N^*$ is the density of excited atoms in the gase. It is seen from expression (\ref{eq27}) at hight concentration of a excited atoms in the gas we have an intense stimulated radiation.

It should be noted that in the case of dense gases had to consider the cooperative effects, and the collision of atoms. These phenomena, we will consider in future studies.

\section{Conclusion}
We consider collision of two atoms with formation of the metastable molecules (Feshbach resonance) and the subsequent stimulated transition under the influence of the laser radiation to a lower unbound molecular electronic state with emission of photon of the laser radiation field. This a situation can develop realized, in particular, for $Rb_2$ molecules due to resonance scattering of two $Rb$ atoms in states $5s$ and $5p$ with formation of metastable molecular state $^3\Pi_u$, with subsequent stimulated transition to the lower $1a^3\Sigma^+_u$ unbound molecular electronic state, with emission of photons of the laser radiation. Expressions for the cross sections of the elastic and inelastic resonance scattering and the intensity of the stimulated emission of photons coming from unity volume of the gas are obtained.

The typical atomic gas density is $10^{14}-10^{17}cm^{-3}$. If the density of excited atoms is $0.001\%$, then the intensity of emitted photons during the simulated transition is $10^9-10^{12}$ times larger. As a result, we have a source of high radiation, which serves as an example of an excimer laser. From the expression (32), the peak of the spectral intensity in the graph at $\omega=E_e+\Delta (\lambda)-\lambda$ is $I(\omega)=N^*I_0\frac{2}{\pi\Gamma}$

\section{Acknowledgments}
We are very grateful to Professor A.V. Papoyan for fruitful discussion. Work was supported by the Ministry of Education and Science of Armenia (MESA) project 15T-1C066

\section{references}

\end{document}